\documentclass{article}
\usepackage{dcase2025_techrep,amsmath,graphicx,url,times,booktabs, tabularx}
\usepackage{multirow}
\usepackage{graphicx}
\newcolumntype{C}{>{\centering\arraybackslash}X}

\title{AISTAT LAB SYSTEM FOR DCASE 2025 TASK6: LANGUAGE-BASED AUDIO RETREIVAL}

\name{Hyun Jun Kim$^{1}$,
      Hyeong Yong Choi$^{2}$,
      Changwon Lim$^{1}\sthanks{Corresponding author}$
      }
\address{$^1$ Department of Smart Cities, Chung-Ang University, Seoul, Korea\\          
        $^2$ Department of Applied Statistics, Chung-Ang Univeristy, Seoul, Korea\\ 
        \{hyunjun0615, chy1218, clim\}@cau.ac.kr\\
 }

\begin{document}

\ninept
\maketitle

\begin{sloppy}

\begin{abstract}
This report presents the AISTAT team’s submission to the language-based audio retrieval task in DCASE 2025 Task 6. Our proposed system employs dual encoder architecture, where audio and text modalities are encoded separately, and their representations are aligned using contrastive learning. Drawing inspiration from methodologies of the previous year’s challenge, we implemented a distillation approach and leveraged large language models (LLMs) for effective data augmentation techniques, including back-translation and LLM mix. Additionally, we incorporated clustering to introduce an auxiliary classification task for further finetuning. Our best single system achieved a mAP@16 of 46.62, while an ensem-ble of four systems reached a mAP@16 of 48.83 on the Clotho development test split. 
\end{abstract}


\section{Introduction}
\label{sec:intro}

DCASE 2025 Task 6 challenge \cite{dcase2025_task6web} focuses on language-based audio retrieval, a task that requires retrieving audio recordings from a database that best matches a given textual query, and vice versa. This task is critical for applications such as con-tent-based multimedia search, audio annotation, and cross-modal understanding, where aligning audio and text modalities in a shared semantic space is essential. Unlike traditional audio classification or tagging, language-based audio retrieval demands models that capture nuanced semantic relationships between free-form text descriptions and complex audio sig-nals, which may contain overlapping or ambiguous acoustic concepts.
Our approach builds on DCASE 2024 Task 8 \cite{dcase2024_task8web}, adopt-ing a dual-encoder architecture with advanced techniques, such as distillation loss, LLM-based data augmentation, and auxiliary classification. These methods aim to enhance the model’s generalization, robustness, and ability to capture fine-grained audio-text relationships.
The remainder of this paper is organized as follows. Sec-tion 2 describes the proposed system in detail. Section 3 out-lines the datasets, models, and training protocols. Finally, Section 4 presents the experimental results and describes the submitted systems.

\section{METHOD}
\label{sec:mthd}

Our system leverages a dual-encoder architecture, where audio and text inputs are processed by separate encoders and aligned in a joint embedding space. We enhance this frame-work with contrastive learning, distillation loss, an auxiliary classification task, and data augmentation, as detailed below. The overall structure is illustrated in Figure 1.

\subsection{Contrastive learning}
\label{ssec:cl}
We employed a contrastive learning framework as the foundational approach to align audio and text representations. Contrastive learning seeks to create a joint embedding space where corresponding audio-text pairs are closely aligned, while non-corresponding pairs are distanced \cite{koepke2022audio}. This is accomplished by optimizing the InfoNCE loss, which maximizes the cosine similarity of matched audio-text embeddings and minimizes it for unmatched pairs within a batch.
Let $\phi_a$ and $\phi_c$ denote the audio and text encoders, respectively, which map audio inputs $a_i$ and text captions $c_j$ to their respective embeddings. The similarity between an audio embedding $\phi_a(a_i)$ and a text embedding $\phi_c(c_j)$ is defined as the normalized cosine similarity:  
\begin{equation}
    C_{ij} = \frac{\phi_a(a_i)^T\cdot\phi_c(c_j)}{\Vert\phi_a(a_i)\Vert_2\Vert\phi_c(c_j)\Vert_2},
\end{equation}
where $\Vert\cdot\Vert_2$ represents the L2 norm, ensuring unit-normalized embeddings. We compute softmax-normalized probabilities for audio-to-text and text-to-audio retrieval as: 

\begin{equation}
    q_a(a_i|c_j)=\frac{exp(C_{ij}/\tau)}{\sum^{N}_{k=1}exp(C_{kj}/\tau)}, 
\end{equation}

\begin{equation}
    q_c(c_j|a_i)=\frac{exp(C_{ij}/\tau)}{\sum^{N}_{l=1}exp(C_{il}/\tau)}, 
\end{equation}

where $\tau>0$ is a temperature parameter controlling the softness of the distribution. We used $\tau=0.05$ in all our experiments. These probabilities reflect the model’s confidence in matching audio $a_i$ to caption $c_j$, and vice versa, relative to other items in the batch. The supervised contrastive loss is the sum of cross-entropy losses between the predicted probabilities $(q_a, q_c)$ and the ground-truth distributions $(p_a,p_c)$, where $p_a$ and $p_c$ assign a probability of 1 to the positive pair and 0 to negative pairs:
\begin{equation}
    L_{sup}=H(p_a,q_a)+H(p_c,q_c),
\end{equation}
where $H$ is the cross-entropy loss.  

\begin{figure*}[t]
  \centering
  \centerline{\includegraphics[width=\textwidth]{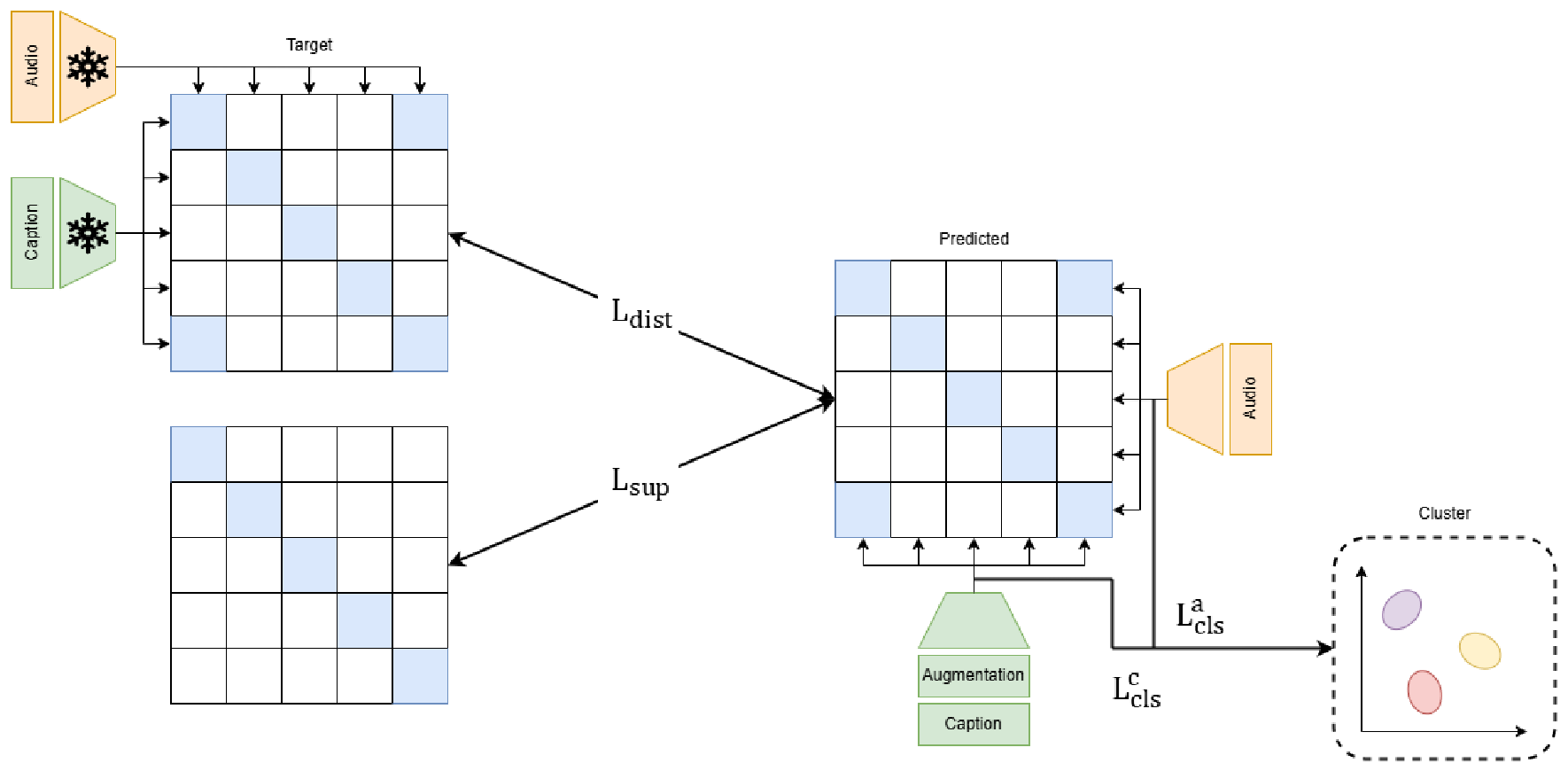}}
  \caption{Overview of our system. The pretrained model is used to generate targets for the distill loss. After the finetuning phase, 
clustering is performed separately on audio and text data to assign pseudo-labels. These pseudo-labels are utilized to introduce an 
auxiliary classification task, which guides the re-finetuning process.}
\end{figure*}

\subsection{Distillation loss}
\label{ssec:dl}

To address the binary correspondence assumption in audio retrieval datasets like ClothoV2, where captions may describe multiple recordings due to overlapping acoustic concepts or limited diversity, we adopted a distillation loss approach from the top-ranked DCASE 2024 Task 8 system \cite{primus2024estimated}. This method uses soft correspondence probabilities from an ensemble of pretrained models to capture nuanced audio-text relationships, improving generalization. 
Formally, we first compute the similarity between audio embedding and text embedding as defined in Section 2.1. An ensemble of $M$ pretrained models generates soft correspondence probabilities by averaging their similarity scores: 
\begin{equation}
    \hat{C}_{ij}=\frac{1}{M}\sum^M_{m=1}C_{ij}^m.
\end{equation}
These averaged similarities are used to compute soft probabilities in a knowledge distillation-like procedure:

\begin{equation}
    \hat{p}_a(a_i|c_j)=\frac{exp(\hat{C}_{ij}/\tau)}{\sum^N_{k=1}exp(\hat{C}_{kj}/\tau)},
\end{equation}

\begin{equation}
    \hat{p}_c(c_j|a_i)=\frac{exp(\hat{C}_{ij}/\tau)}{\sum^N_{l=1}exp(\hat{C}_{il}/\tau)},
\end{equation}

The distillation loss is calculated as the cross-entropy between these soft probabilities and the model’s predicted probability:
\begin{equation}
    L_{dist}=H(\hat{p}_a,q_a)+H(\hat{p}_c,q_c).
\end{equation}
The total loss combines the supervised contrastive loss $L_{sup}$  with the distillation loss, weighted by $\lambda=1.0$:
\begin{equation}
    L=L_{sup}+\lambda L_{dist}
\end{equation}
By leveraging these soft targets, the distillation loss enhances the model’s ability to capture complex relationships between audio recordings and captions, improving its generalization across diverse audio-text pairs.

\begin{table}[t]
  \centering
  \caption{System ID (SID) for various training configurations}
  \vspace{1pt}
  \begin{tabular}{cccc}
    \hline
    SID & Distill & Augmentation & Cluster weight \\
    \hline
    1 & X & X & X \\
    2 & O & X & X \\
    3 & O & O & X \\
    4 & O & O & Finetuned \\
    5 & O & O & BERTopic \\
    \hline
  \end{tabular}
\end{table}

\subsection{Cluster-based classification}
\label{ssec:cbc}

We propose a novel approach to enhance language-based audio retrieval by introducing an auxiliary classification task to further improve the model's representation learning. We perform clustering on all captions in the Clotho dataset to lay the foundation for an auxiliary task. We generate embedding for each caption and apply a clustering method similar to BERTopic \cite{grootendorst2022bertopic}, which typically involves dimensionality reduction, such as UMAP \cite{mcinnes2018umap}, followed by density-based clustering, such as HDBSCAN \cite{mcinnes2017hdbscan}, to group captions into semantically similar clusters. Each caption is thus assigned to a specific cluster, representing latent topics or semantic patterns within the captions. 

\begin{table*}[!hbt]
\centering
\begin{tabular}{wc{1.2cm}wc{2cm}wc{1.8cm}wc{1.8cm}wc{1.8cm}wc{1.8cm}wc{1.8cm}wc{1.8cm}}
\hline
\multirow{2}{*}{SID} & \multirow{2}{*}{Audio model} & \multicolumn{2}{c}{Multiple   annotation} & \multicolumn{4}{c}{Single   annotation} \\ \cline{3-8} 
 &  & mAP@10 & mAP@16 & mAP@10 & R@1 & R@5 & R@10 \\ \hline
\multirow{3}{*}{1} & PaSST & 39.45 & 42.08 & 35.47 & 23.35 & 52.5 & 65.07 \\
 & EAT & 38.11 & 40.41 & 35.13 & 23.44 & 51.12 & 63.87 \\
 & BEATs & 35.66 & 38.12 & 34.15 & 22.74 & 49.51 & 63.75 \\ \hline
\multirow{3}{*}{2} & PaSST & 43.75 & \textbf{46.62} & 39.32 & 26.81 & 56.61 & 70.07 \\
 & EAT & 42.83 & 45.35 & 39.50 & 26.79 & 56.40 & 69.44 \\
 & BEATs & 41.36 & 43.89 & 37.92 & 25.26 & 54.81 & 69.00 \\ \hline
\multirow{3}{*}{3} & PaSST & 43.56 & 46.41 & 39.92 & 27.20 & \textbf{57.84} & 70.74 \\
 & EAT & 43.37 & 46.05 & \textbf{40.28} & \textbf{27.52} & 57.63 & \textbf{71.35} \\
 & BEATs & 42.09 & 44.66 & 38.42 & 25.51 & 56.02 & 69.44 \\ \hline
\multirow{3}{*}{4} & PaSST & 43.61 & 46.39 & 39.92 & 27.2 & 57.21 & 70.24 \\
 & EAT & 42.83 & 45.34 & 40.02 & 27.43 & 56.59 & 70.62 \\
 & BEATs & 42.01 & 44.58 & 38.61 & 25.88 & 55.94 & 69.46 \\ \hline
\multirow{3}{*}{5} & PaSST & \textbf{43.79} & 46.50 & 39.58 & 26.66 & 57.38 & 70.14 \\
 & EAT & 42.65 & 45.34 & 39.73 & 26.67 & 57.28 & 70.18 \\
 & BEATs & 41.32 & 43.88 & 38.23 & 25.26 & 56.06 & 69.86 \\ \hline
\multicolumn{8}{c}{Ensemble} \\ \hline
\multicolumn{2}{c}{E1} & \textbf{46.07} & \textbf{48.83} & 41.60 & 28.33 & 59.71 & 72.06 \\
\multicolumn{2}{c}{E2} & 46.05 & 48.78 & 41.58 & 28.34 & 59.87 & 72.23 \\
\multicolumn{2}{c}{E3} & 46.03 & 48.80 & 41.70 & \textbf{28.46} & 59.85 & 72.38 \\
\multicolumn{2}{c}{E4} & 46.04 & 48.79 & \textbf{41.72} & 28.38 & \textbf{60.02} & \textbf{72.46} \\ \hline
\end{tabular}%
\caption{Retrieval performance of the models (first section) and the ensembled systems (second section). Note that SID stands for System ID, which is detailed in Table 1.}
\label{tab:my-table}
\end{table*}

To leverage the clustering results, we extend the model architecture by adding classification heads to both the text and audio encoders. The classification head for the text encoder is designed to predict the cluster label of the input caption, while the audio encoder’s classification head predicts the cluster label of the corresponding caption. Specifically, the output of each encoder is processed through two sequential linear layers with a ReLU activation function between them, projecting the output to a vector with dimensions equal to the number of clusters. The intermediate linear layer has a dimension three times that of the input to enhance representation capacity. This setup encourages the audio encoder to learn representations that are aligned with the semantic clusters of the captions, thereby enhancing the fine-grained alignment between audio and text. 
The total loss combines the supervised contrastive loss $L_{sup}$  from Section 2.1, the distillation loss $L_{dist}$ from Section 2.2, and the classification losses for the audio and text encoders, denoted  $L_{cls}^a$ and $L_{cls}^c$, respectively:
\begin{equation}
    L=L_{sup}+\lambda_1L_{dist}+\lambda_2(L_{cls}^a+L_{cls}^c)
\end{equation}
In all experiments, we fixed $\lambda_1=1.0$ and $\lambda_2=0.05$ to balance the contributions of each loss term.

\subsection{Data augmentation}
\label{ssec:da}

To enhance the diversity of captions for our text-grounded audio retrieval, we employed caption augmentation leveraging the capabilities of a large language model (LLM), specifically GPT-4o \cite{hurst2024gpt}. One of the key techniques utilized was \textbf{back-translation} \cite{sennrich2015improving}. This method involves translating the original English captions into a randomly selected language and then translating them back into English. By doing so, back-translation generates captions that retain the same semantic meaning as the originals but feature varied linguistic expressions. 
In addition to back-translation, we implemented another augmentation technique called \textbf{LLM mix} \cite{wu2024improving} to further enrich our dataset. For this method, we randomly selected two audio-text pairs and combined their audio signals to create a new mixed audio sample. To generate a corresponding caption for this mixed audio, we utilized GPT-4o to intelligently merge the captions of the original audio-text pairs. With LLM mix, we created 50,000 new audio-caption pairs, adding substantial variety to our dataset

\section{Experiments}
\label{sec:exp}

The following subsections provide comprehensive details on the datasets, models, and training protocols to ensure repro-ducibility.

\subsection{Datasets}
\label{ssec:data}

\textbf{ClothoV2.1} \cite{drossos2020clotho} comprises audio recordings with durations ranging from 15 to 30 seconds, each accompanied by captions containing 8 to 20 words. The development set is divided into training, validation, and test splits. Each recording is paired with five captions created by human annotators. 

\textbf{AudioCaps} \cite{kim2019audiocaps} consists of 51,308 audio recordings sourced from AudioSet, each 10 seconds long and paired with a single human-generated caption. The captions have an average length of 9.8 words. For our experiments, we combined the training, validation, and test splits of AudioCaps into a single dataset, which was used for pretraining the model.

\textbf{WavCaps} \cite{mei2024wavcaps} is a weakly-labeled dataset containing 403,050 audio recordings of varying durations, collected from sources including FreeSound, BBC Sound Effects, SoundBible, and the strongly supervised subset of AudioSet. To adhere to this year’s updated competition rules, we excluded any recordings in WavCaps that overlapped with the evaluation subsets of ClothoV2 and were used for pretraining as well.

\begin{table*}[!hbt]
\centering
\begin{tabular}{ccccccccccccc}
\hline
SID & \multicolumn{3}{c}{2} & \multicolumn{3}{c}{3} & \multicolumn{3}{c}{4} & \multicolumn{3}{c}{5} \\
Model & PaSST & EAT & BEATs & PaSST & EAT & BEATs & PaSST & EAT & BEATs & PaSST & EAT & BEATs \\ \hline
E1 & 0.2275 & 0.07 & 0.06 & 0 & 0.12 & 0.045 & 0.325 & 0 & 0.045 & 0.0975 & 0.01 & 0 \\
E2 & 0.2275 & 0.0875 & 0.04 & 0 & 0.15 & 0.03 & 0.325 & 0 & 0.03 & 0.0975 & 0.0125 & 0 \\
E3 & 0.225 & 0.175 & 0.1 & 0.03 & 0.01 & 0.01 & 0.195 & 0.045 & 0.06 & 0.09 & 0.03 & 0.03 \\
E4 & 0.18 & 0.14 & 0.08 & 0.09 & 0.03 & 0.03 & 0.13 & 0.03 & 0.04 & 0.15 & 0.05 & 0.05 \\ \hline
\end{tabular}%
\caption{Combination coefficients for four submitted system}
\label{tab:my-table}
\end{table*}

\subsection{Audio embedding models}
\label{ssec:audioenc}

\textbf{The Patchout faSt Spectrogram Transformer (PaSST)} \cite{koutini2021efficient} leverages pre-trained parameters from a vision transformer and fine-tunes them on the AudioSet dataset for general-purpose audio tagging. By dropping patches from the input sequence, PaSST achieves a low computational and memory footprint. In our experiments, we used a PaSST version without patch overlap, applying structured patchout of 2 and 15 over the frequency and time dimensions, respectively.

\textbf{The Efficient Audio Transformer (EAT)} \cite{chen2024eat} is an audio self-supervised learning (SSL) model focused on efficient representation learning from unlabeled audio data. It employs a novel Utterance-Frame Objective (UFO) that combines global utterance-level and local frame-level learning to improve audio understanding. We initialized the models with publicly available pretrained weights, namely EAT-base\_epoch30\_pt. 

\textbf{Bidirectional Encoder representation from Audio Transformers (BEATs)} \cite{chen2022beats} is a self-supervised learning framework designed for pre-training comprehensive audio representations. It integrates an acoustic tokenizer with an audio SSL model, optimized iteratively to generate discrete labels rich in audio semantics. We also initialized BEATs with publicly available pretrained weights, namely BEATs\_iter3\_plus\_AS2M.

\subsection{Sentence embedding models}
\textbf{RoBERTa} \cite{liu2019roberta} is a BERT-based language model developed by Facebook AI that improves upon the original BERT pre-training methodology. By removing the Next Sentence Prediction (NSP) objective, extending training duration, increasing batch size, and leveraging a larger and more diverse corpus, RoBERTa achieves stronger performance in sentence-level representation learning. In our experiments, we used RoBERTa-large as a sentence embedding extractor, utilizing its pretrained parameters to capture rich semantic information from textual inputs.

\subsection{Training}

Audio inputs were preprocessed to align with the pretraining configurations of the respective models. Specifically, EAT and BEATs used a sampling rate of 16 kHz, while PaSST used 32 kHz. In all cases, audio was converted to log-mel spectrograms as the input representation. All models were trained using the AdamW optimizer. Learning rates were adjusted using a cosine warmup scheduler, with specific values detailed in the respective training stages. The training process was divided into three stages. Initial pretraining was conducted on the CLOTHO, WavCaps, and Audi
oCaps datasets to learn general audio-text alignment, while the subsequent finetuning and re-finetuning stages were performed exclusively on the CLOTHO dataset. Each stage is described below. 
 
\textbf{Initial pretraining} – We use a mix of Clotho development training split, AudioCaps, and WavCaps datasets. The training spans 20 epochs. No data augmentation is applied in this phase. Due to computational resource constraints, we set batch size to 64 for PaSST, 24 for EAT, and 16 for BEATs. To accommodate these configurations, we adjusted the learning rates using a cosine warmup scheduler across all training processes. For PaSST, the 
learning rate decreased from 2e-5 to 1e-7, while for EAT and BEATs, it decreased from 1e-5 to 1e-7. These hyperparameter settings were consistently applied in the subsequent finetuning and re-finetuning stages. 
 
\textbf{Finetuning} – In the finetuning phase, models were further trained for 20 epochs using ensemble soft labels. Soft labels were calcu
lated as the average of the similarity matrices obtained from three audio models, as described in Equation 5, where M equals 3. 
These soft labels served as targets for a distillation loss, guiding the model toward a consensus representation. To enhance robust
ness, we trained models with and without data augmentation. For the augmented models, we applied data augmentation techniques, 
including back-translation and LLM-based caption mixing, as described in section 2.4. Random deletion and synonym replace
ment were also applied to a single word in captions with an 80\% probability, further increasing caption diversity.  
 
\textbf{Re-finetuning with cluster-guided classification} – In the refinetuning phase, we enhanced our model through cluster-guided classification. Clustering was conducted using two weight sources: our finetuned model weights and the pre-trained e5-large-v2 weights, sourced from the e5 model family and utilized within the BERTopic framework [5, 18]. The e5-large-v2 model excels in clustering tasks by generating high-quality sentence embeddings that preserve semantic similarity in the embedding space. For each embedding set, we employed the BERTopic framework with 
HDBSCAN to assign pseudo-labels to text samples, reassigning outliers based on topic probabilities estimated by BERTopic. Refinetuning spanned 20 epochs.  
 
We evaluated four systems combining pretraining, distillation, caption augmentation, and cluster supervision. The configuration of these variants is summarized in Table 1.  

\section{Results}

Table 2 presents the performance of our four systems on the ClothoV2 development test split. The systems, detailed in Table 1, vary in their use of distillation, data augmentation, and clustering, with three audio models. PaSST consistently outperformed EAT and BEATs across all systems, achieving the highest mAP@16.  A weighted ensemble of Systems 2–5 significantly improved performance over individual systems. We employed two ensem
ble strategies. In methods E1 and E2, we first calculated system level ensembles across Systems 2–5 and then computed weighted sums for each model. Conversely, in methods E3 and E4, we first computed model-level ensembles for each model by combining outputs from Systems 2–5, then performed a weighted sum across the systems. The weights for all ensembles were determined through grid search to optimize mAP@16 on the validation set. By leveraging the complementary strengths of the systems and models, the ensembles achieved a highest mAP@16 of 48.83. 
For the final submission, we retrained all systems on the entire development split of the ClothoV2 dataset and submitted the weighted sum of their similarity matrices using the weights from Table 3.

\section{Conclusion}

This paper described the AISTAT Lab’s system for text-grounded audio retrieval system. Inspired by the methodologies of top-performing teams in the previous year, we applied data augmentation techniques leveraging LLMs and incorporated a distillation loss to enhance our model’s performance. Furthermore, by utilizing clustering, we introduced an auxiliary classification task to the training process, which contributed to additional performance gains. These combined strategies enabled our system to achieve improved results. 

\section{Acknowledgment}
This work was supported by the National Research Founda-tion of Korea(NRF) grant funded by the Korea govern-ment(MSIT)(RS-2024-00360176).

\bibliographystyle{IEEEtran}
\bibliography{refs}

\begin{thebibliography}{10}
\providecommand{\url}[1]{#1}
\def\UrlFont{\rmfamily}
\providecommand{\newblock}{\relax}
\providecommand{\bibinfo}[2]{#2}
\providecommand\BIBentrySTDinterwordspacing{\spaceskip=0pt\relax}
\providecommand\BIBentryALTinterwordstretchfactor{4}
\providecommand\BIBentryALTinterwordspacing{\spaceskip=\fontdimen2\font plus
\BIBentryALTinterwordstretchfactor\fontdimen3\font minus \fontdimen4\font\relax}
\providecommand\BIBforeignlanguage[2]{{%
\expandafter\ifx\csname l@#1\endcsname\relax
\typeout{** WARNING: IEEEtran.bst: No hyphenation pattern has been}%
\typeout{** loaded for the language `#1'. Using the pattern for}%
\typeout{** the default language instead.}%
\else
\language=\csname l@#1\endcsname
\fi
#2}}

\bibitem{dcase2025_task6web}
\url{https://dcase.community/challenge2025/task-language-based-audio-retrieval}.

\bibitem{dcase2024_task8web}
\url{https://dcase.community/challenge2024/task-language-based-audio-retrieval}.

\bibitem{koepke2022audio}
A.~S. Koepke, A.-M. Oncescu, J.~F. Henriques, Z.~Akata, and S.~Albanie, ``Audio retrieval with natural language queries: A benchmark study,'' \emph{IEEE Transactions on Multimedia}, vol.~25, pp. 2675--2685, 2022.

\bibitem{primus2024estimated}
P.~Primus, F.~Schmid, and G.~Widmer, ``Estimated audio-caption correspondences improve language-based audio retrieval,'' \emph{arXiv preprint arXiv:2408.11641}, 2024.

\bibitem{grootendorst2022bertopic}
M.~Grootendorst, ``Bertopic: Neural topic modeling with a class-based tf-idf procedure,'' \emph{arXiv preprint arXiv:2203.05794}, 2022.

\bibitem{mcinnes2018umap}
L.~McInnes, J.~Healy, and J.~Melville, ``Umap: Uniform manifold approximation and projection for dimension reduction,'' \emph{arXiv preprint arXiv:1802.03426}, 2018.

\bibitem{mcinnes2017hdbscan}
L.~McInnes, J.~Healy, S.~Astels, \emph{et~al.}, ``hdbscan: Hierarchical density based clustering.'' \emph{J. Open Source Softw.}, vol.~2, no.~11, p. 205, 2017.

\bibitem{hurst2024gpt}
A.~Hurst, A.~Lerer, A.~P. Goucher, A.~Perelman, A.~Ramesh, A.~Clark, A.~Ostrow, A.~Welihinda, A.~Hayes, A.~Radford, \emph{et~al.}, ``Gpt-4o system card,'' \emph{arXiv preprint arXiv:2410.21276}, 2024.

\bibitem{sennrich2015improving}
R.~Sennrich, B.~Haddow, and A.~Birch, ``Improving neural machine translation models with monolingual data,'' \emph{arXiv preprint arXiv:1511.06709}, 2015.

\bibitem{wu2024improving}
S.-L. Wu, X.~Chang, G.~Wichern, J.-w. Jung, F.~Germain, J.~Le~Roux, and S.~Watanabe, ``Improving audio captioning models with fine-grained audio features, text embedding supervision, and llm mix-up augmentation,'' in \emph{ICASSP 2024-2024 IEEE International Conference on Acoustics, Speech and Signal Processing (ICASSP)}.\hskip 1em plus 0.5em minus 0.4em\relax IEEE, 2024, pp. 316--320.

\bibitem{drossos2020clotho}
K.~Drossos, S.~Lipping, and T.~Virtanen, ``Clotho: An audio captioning dataset,'' in \emph{ICASSP 2020-2020 IEEE International Conference on Acoustics, Speech and Signal Processing (ICASSP)}.\hskip 1em plus 0.5em minus 0.4em\relax IEEE, 2020, pp. 736--740.

\bibitem{kim2019audiocaps}
C.~D. Kim, B.~Kim, H.~Lee, and G.~Kim, ``Audiocaps: Generating captions for audios in the wild,'' in \emph{Proceedings of the 2019 Conference of the North American Chapter of the Association for Computational Linguistics: Human Language Technologies, Volume 1 (Long and Short Papers)}, 2019, pp. 119--132.

\bibitem{mei2024wavcaps}
X.~Mei, C.~Meng, H.~Liu, Q.~Kong, T.~Ko, C.~Zhao, M.~D. Plumbley, Y.~Zou, and W.~Wang, ``Wavcaps: A chatgpt-assisted weakly-labelled audio captioning dataset for audio-language multimodal research,'' \emph{IEEE/ACM Transactions on Audio, Speech, and Language Processing}, vol.~32, pp. 3339--3354, 2024.

\bibitem{koutini2021efficient}
K.~Koutini, J.~Schl{\"u}ter, H.~Eghbal-Zadeh, and G.~Widmer, ``Efficient training of audio transformers with patchout,'' \emph{arXiv preprint arXiv:2110.05069}, 2021.

\bibitem{chen2024eat}
W.~Chen, Y.~Liang, Z.~Ma, Z.~Zheng, and X.~Chen, ``Eat: Self-supervised pre-training with efficient audio transformer,'' \emph{arXiv preprint arXiv:2401.03497}, 2024.

\bibitem{chen2022beats}
S.~Chen, Y.~Wu, C.~Wang, S.~Liu, D.~Tompkins, Z.~Chen, and F.~Wei, ``Beats: Audio pre-training with acoustic tokenizers,'' \emph{arXiv preprint arXiv:2212.09058}, 2022.

\bibitem{liu2019roberta}
Y.~Liu, M.~Ott, N.~Goyal, J.~Du, M.~Joshi, D.~Chen, O.~Levy, M.~Lewis, L.~Zettlemoyer, and V.~Stoyanov, ``Roberta: A robustly optimized bert pretraining approach,'' \emph{arXiv preprint arXiv:1907.11692}, 2019.

\end{thebibliography}

\end{sloppy}
\end{document}